\documentclass[10pt,a4paper,twocolumn,english,aps,manuscript,superscriptaddress]{revtex4}

\usepackage[T1]{fontenc}
\usepackage[latin9]{inputenc}
\usepackage{color}
\usepackage{babel}
\usepackage{latexsym}
\usepackage{float}
\usepackage{amsmath}
\usepackage{graphicx}
\usepackage{times}   %% Times Roman font
\usepackage{esint}
\usepackage[unicode=true,pdfusetitle,
 bookmarks=false,colorlinks=true,citecolor=blue,urlcolor=blue,linkcolor=red]{hyperref}

\makeatletter

\@ifundefined{textcolor}{}
{%
 \definecolor{BLACK}{gray}{0}
 \definecolor{WHITE}{gray}{1}
 \definecolor{RED}{rgb}{1,0,0}
 \definecolor{GREEN}{rgb}{0,1,0}
 \definecolor{BLUE}{rgb}{0,0,1}
 \definecolor{CYAN}{cmyk}{1,0,0,0}
 \definecolor{MAGENTA}{cmyk}{0,1,0,0}
 \definecolor{YELLOW}{cmyk}{0,0,1,0}
}

\@ifundefined{date}{}{\date{}}
\AtBeginDocument{
  
}
\makeatother

\newcommand{\SAVE}[1]{}

\newcommand{\prlsec}[1]{\emph{#1---}}
\newcommand{\QD}{{\rm QD}}
\newcommand{\Ncal}{{\mathcal N}}
\newcommand{\nhalf}{{n_{1/2}}}
\newcommand{\nspinone}{{n_1}}
\newcommand{\Ttilde}{{\widetilde{\mathbf{T}}}}
\newcommand{\T}{{\mathbf{T}}}
\newcommand{\phitilde}{{\widetilde{\phi}}}
\newcommand{\DLOW}{{\Delta_{low}}}
\newcommand{\DQD}{{\Delta}}
\newcommand{\SQD}{{\sigma}_{\QD}}
\newcommand{\SIQD}{\mathbf{S}_{i,\QD}}

\newcommand{\Heff}{\mathcal{H}_{\rm eff}}

\begin{document}
\renewcommand\abstractname{}

\title{Emergent spin excitations in a Bethe lattice at percolation}
\author{Hitesh J. Changlani}
\thanks{Present address: Department of Physics, University of Illinois at Urbana-Champaign, Urbana, Illinois 61801, USA}
\affiliation{Laboratory of Atomic And Solid State Physics, Cornell University, Ithaca, NY 14853, USA}
\author{Shivam Ghosh}
\affiliation{Laboratory of Atomic And Solid State Physics, Cornell University, Ithaca, NY 14853, USA}
\author{Sumiran Pujari}
\affiliation{Laboratoire de Physique Th\'{e}orique, Universit\'{e} Paul Sabatier, 31062 Toulouse, France}
\author{Christopher L. Henley}
\affiliation{Laboratory of Atomic And Solid State Physics, Cornell University, Ithaca, NY 14853, USA}
\date{October 7, 2013}

\begin{abstract}
We study the spin 1/2 quantum Heisenberg antiferromagnet on a Bethe
lattice diluted to the percolation threshold. Dilution creates areas of
even/odd sublattice imbalance resulting in ``dangling spins''
(L. Wang and A. W. Sandvik, Phys. Rev. Lett. 97, 117204 (2006); Phys.
Rev. B 81, 054417 (2010)). These collectively act as ``emergent'' spin 1/2 degrees of freedom and 
are responsible for the creation of a set of low lying ``quasi degenerate states''. 
Using density matrix renormalization group (DMRG) calculations, 
we detect the presence and location of these emergent spins. 
We find an effective Hamiltonian of these emergent spins, 
with Heisenberg interactions that decay \emph{exponentially} with the distance between them.
\end{abstract}

\maketitle

\prlsec{Introduction}Quantum spins on percolation clusters~\cite{aharony}
provide an ideal testbed for studying the interplay between 
geometrical disorder and quantum fluctuations. 
The Hamiltonian for these problems is
\begin{equation}
\mathcal{H}=\sum_{\langle ij\rangle}J{\mathbf{S}_{i}\cdot{\mathbf{S}_{j}
\label{eq:hamiltonian}}}
\end{equation}
where ${\bf {S}_{i}}$ are Pauli spin $1/2$ operators and the sum runs over nearest-neighbor 
occupied sites, and $J>0$.
Theoretical~\cite{chen,bray-Ali,sandvik2002,Yu} and experimental~\cite{vajk} 
studies of quantum spins on diluted square lattices have 
focused on the question of whether long range order survives up to
the classical percolation threshold $p_{c}$. 
A numerical study~\cite{sandvik2002}
%seems to have 
has settled this question and found long range order to be robust to quantum
fluctuations, surviving all the way up to $p_{c}$.

The excitations are less straightforward.
For uniform lattices with number of sites $N$, the lowest energy scale 
{\sl consistent with N\'eel order breaking a continuous symmetry} is $\sim JN^{-1}$, 
corresponding to a "tower" of states: mixtures of symmetry-broken states that 
become degenerate in the thermodynamic limit~\cite{Anderson,Ziman,Gross}.
However a Quantum Monte Carlo study by Wang and Sandvik~\cite{wangandsandvik} discovered
a somewhat "anomalous" finite size scaling
of the spin gap $\DLOW$: $\DLOW\approx N^{-2}$ (for clusters with a singlet ground state)
or $\DLOW \approx N^{-1.5}$ (for generic clusters, most with a non-singlet ground state). 
A strong correspondence was shown~\cite{wangandsandvik} between these low lying states 
and places on the cluster where there is a local imbalance between the
number of even and odd sites.  It was conjectured that, in each such place, 
a spin degree emerges which is effectively decoupled from the antiferromagnetic
order and hence was called a ``dangling spin.''
%In this context, we remark that the effect of quenched dilution in the background of classical spins on a 
%two-dimensional lattice has been studied extensively by~\cite{Harris_Kirkpatrick}, however, 
%the effects we describe here require a full quantum mechanical treatment. 

\begin{figure}[htpb]
\centering
 \includegraphics[width=\linewidth]{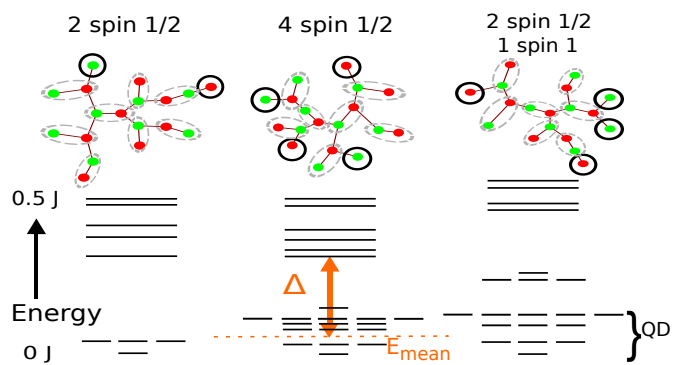} \caption{(Color online) 
Three different percolation clusters (all of the same size $N=18$) are shown with
their corresponding low energy spectra. The red (dark) and green (light) circles
indicate even and odd sites. 
The broken dashed lines show dimer coverings which serve as a heuristic to locate the "dangling spins" (circled with thick black lines). 
Energy spectra for each of the clusters show low lying quasi degenerate (QD) states separated from the continuum by an energy scale $\DQD$. 
$\SQD$ (not shown) is a measure of the spread of QD energies.
}
\label{fig:qd} 
\end{figure}

The goal of this Letter is to characterize the dangling-spin degrees
of freedom numerically,
relating their nature to the local geometry of the cluster, and
to explain the observed low energy spectrum in terms of mediated interactions
between dangling spins.
Our Hamiltonian is \eqref{eq:hamiltonian}
on clusters obtained by randomly diluting the Bethe lattice of coordination 3
at its percolation threshold,
$p_{c}=1/2$ (see examples of small clusters in Fig.~\ref{fig:qd}).
The lack of loops in the Bethe lattice is conducive for using the 
Density Matrix Renormalization Group (DMRG)~\cite{White_PRL}
algorithm, as adapted to generic tree graphs~\cite{pure_bethe}, 
to obtain ground and (some) excited states.

In the rest of this Letter, we first show that 
a typical percolation cluster's spectrum has a clearly separated
low energy component, with a multiplicity consistent with the
expected number of weakly coupled spin-1/2 (sometimes spin-1) dangling spins.
We next show that each dangling spin is somewhat delocalized over a few sites:
on the one hand, we model it as an unpaired spin in a dimerized background
to predict the dangling spin's nature from the local geometry; on
the other hand, by processing spin expectations we obtain the explicit
``localized state wavefunction'' for each dangling spin.
Finally, for each cluster we construct the effective Hamiltonian of the
emergent dangling spins, consisting of \emph{pairwise}, unfrustrated exchange 
interactions decaying exponentially with separation, 
mediated by the background of almost dimerized spins on the balanced parts of the
cluster;  this accurately reproduces the details of that
cluster's low-energy spectrum.

\prlsec{Exact correspondence between dangling spins and low energy spectrum}
We carried out DMRG calculations for several hundred balanced clusters (i.e. having equal number of even and odd sublattice sites)~\cite{FN-balanced}
for sizes up to $N=100$, targeting multiple excited states in the low energy spectrum. 
Since the number of low energy states was found to increase rapidly with 
an increase in the number of dangling spins, we restricted our analysis to the case of four dangling spins~\cite{FN-sixdangling}.
%For analyses where only the ground state properties were of interest or where only the lowest energy gap $\DLOW$ was required, 
%larger clusters ($N=200$), with an \emph{arbitrary} number of dangling spins, were also studied. 

In a typical percolation cluster, we observed
a distinct set of low-lying energy levels we shall call "quasi-degenerate" (QD)~\cite{Lhuillier} 
since (we claim) they would be exactly degenerate in the limit that the dangling spins are fully decoupled from the rest of the sites.
The QD states are separated from the continuum of higher energy 
%Quantum Rotor and spin wave 
states by a finite size gap we call $\DQD$ (specifically defined as
the difference between the mean of QD levels and the lowest non-QD
level). The set of QD states are identified by looking at the difference in energies of consecutive states 
up to the Quantum rotor excitation and finding the pair of states with the largest gap. 
The lower energy state in this pair and all states below that make up the QD spectrum~\cite{FN-QD}. 
The energy scale characterizing the spread of the QD states, $\SQD$, is defined to be the standard deviation of the QD energies from their mean value. 
The ratio $r=\langle \SQD/\DQD \rangle$ (where $\langle ...\rangle$ indicates an average over disorder) was found to be small (for example $r = 0.17 \pm 0.1$ for $N=50$), 
justifying our notion of a separation of scales.

Fig.~\ref{fig:qd} also shows a striking correspondence
between the number of low lying QD states, $\Ncal_\QD$, and the
number of dangling spins $n_{d}$ on the percolation cluster.
We find that $\Ncal_\QD=2^\nhalf 3^\nspinone$, where $\nhalf$ and $\nspinone$
are integers and $\nhalf+2\nspinone=n_{d}$. 
Our interpretation of this multiplicity is that $2\nspinone$ of the 
dangling spins pair up so as to form a spin-1, while the others 
remain as spin-half degrees of freedom.
There is thus a one-to-one correspondence between the low-energy 
(QD) eigenstates and the Hilbert space of the posited emergent spins. 
We used an algorithm (described later) that relies only on the cluster geometry
to objectively predict the numbers $\nhalf$ and $\nspinone$ for each cluster,
and verified that their predicted relationship with $\Ncal_\QD$ was satisfied
in every cluster.

We also directly measured the lowest singlet-triplet gaps $\DLOW$ 
for an ensemble of balanced clusters (for sizes up to $N=200$ and not constraining the number of dangling spins). 
Its typical value scales as $N^{-1.9 \pm 0.1}$, which appears remarkably similar to the scaling previously seen on 
square lattice percolation clusters~\cite{wangandsandvik}. 

\prlsec{Locating Dangling degrees of freedom in real space}
Having established the presence of emergent spin-half and spin-one 
degrees of freedom,
we now develop two complementary ways of looking at them.

The first is within the framework of a quantum monomer-dimer model. 
We imagine that the wavefunction is a product of
valence bonds in which the $N$ spins are paired (dimerized)
into singlets to the maximum extent possible (optimal
configuration).
Even when even and odd sites are balanced \emph{globally},
there remain some uncovered sites, i.e. monomers,  
due to \emph{local} imbalances.
These are spin-1/2 degrees of freedom and 
(within this picture) represent the dangling spins.
There are multiple ways to optimally place the monomers; 
the actual wavefunction is imagined to be a superposition of these ways.

Our geometric algorithm, based on the valence bond framework, finds one element from the set 
of optimal dimerizations of the cluster and 
then attempts to find other elements of the set by locally replacing monomers with adjacent dimers. 
In spirit, this is a ``greedy'' algorithm which tries to place dimers wherever possible (to obtain an optimal dimerization pattern),
working from the outer sites inwards on the cluster.

\begin{figure}[htpb]
\centering
\includegraphics[width=1\linewidth]{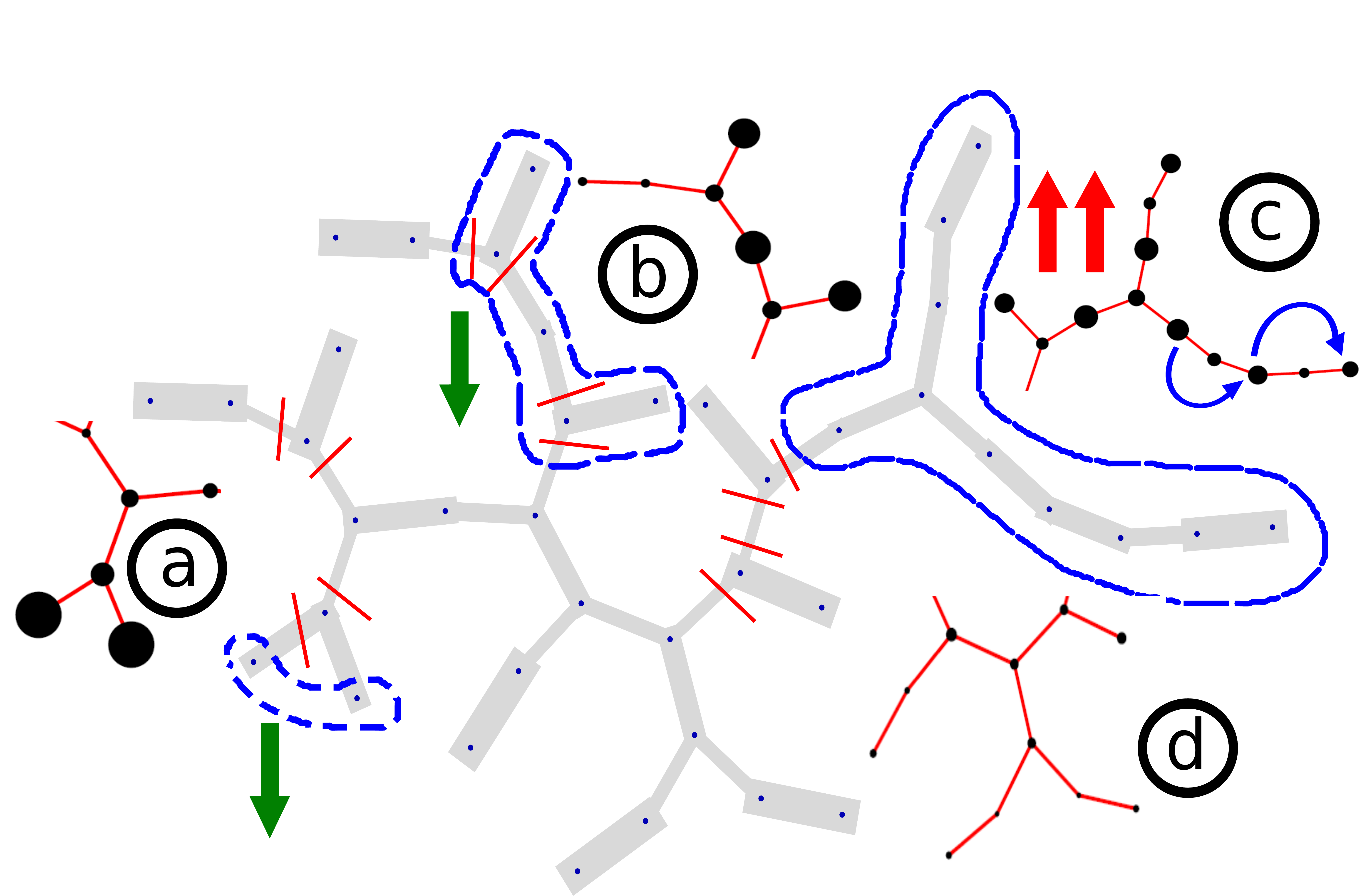}
\caption{(Color online):Typical geometrical motifs in Cayley 
tree percolation clusters, as related to monomer/dimer construction. 
We disconnect the cluster into ``spans'' at the ``prong'' bonds, as indicated by the (red) cut lines. 
The (blue) dashed loops indicate regions with non zero monomer density. 
The thickness of the (grey) bonds is directly proportional 
to $\langle \mathbf{S}_i\cdot \mathbf{S}_j \rangle$ (with the thickest bonds having a value of $\approx $ -0.67). Spatial profiles associated with ``dangling spins", 
are shown on the subclusters marked (a)-(d). The area of the black circles is proportional to 
$\omega_{ii}$ given by Eq.~\eqref{eq:omega_ij}. (a) shows a ``fork" (b) shows a site surrounded by two ``prongs" (c) shows a subcluster where two monomers on the 
same sublattice are present forming an effective spin-1. The (blue) arrows indicate the monomer is free to 
hop around (delocalize) within the subcluster.
(d) shows a region where the spins are ``inert" (largely dimerized).}
%and do not participate in the low energy physics.}
\label{fig:geom} 
\end{figure}	

Given any cluster, there are two operations which cut it down 
to a smaller cluster or clusters, such that all optimal dimerizations
on the smaller cluster(s) are in 1-to-1 correspondence with {\it some} 
of the dimerizations on the larger one.   
The first operation is that, wherever two sites have
coordination 1 or 2, we can remove both (given the dimerization
on the smaller cluster, just insert another dimer to get the 
dimerization on the larger one).  The second operation is that,
wherever we find a pair of adjacent sites with respective 
coordinations 3 and 1 (a ``prong''),
we can always place a dimer on that pair, which fragments the
rest into two subclusters (Fig.~\ref{fig:geom}); a very common special case 
is the fork (Fig.~\ref{fig:geom}(a)), at which we can arbitrarily choose either
side to be the ``prong''.  These two operations can be used
recursively till only isolated sites remain, each 
corresponding to one monomer in the original cluster. Furthermore, any other optimal dimerization is accessible 
from the special one we constructed, by repeatedly
exchanging a monomer with an adjacent dimer.

A monomer can thus "hop" to sites on the lattice via such 
local monomer-dimer rearrangements as shown in Fig.~\ref{fig:geom}(c). 
Our rule-of-thumb is that two monomers (of the same sublattice)
form a spin-1 if and only if they can approach to the minimal
separation of two steps~\cite{FN-spin1}.

%Fig.~\ref{fig:geom} shows some typical geometrical motifs seen 
%in an ensemble of percolation clusters. 
%In Fig.~\ref{fig:geom}(a), a spin-half dangling 
%spin is localized at the tip of the fork. The arms of the fork have a strong dimerization
%and prevent the localized monomer from hopping to other sites in the cluster. Fig.~\ref{fig:geom}(b) shows 
%another type of spin-half degree of freedom,
%which is trapped in the interval between (or on) two ``prongs''. Fig.~\ref{fig:geom}(c) is an example of an emergent spin-one excitation. 
%For comparison, we also show in Fig.~\ref{fig:geom}(d) a region of strong dimerization with no dangling spins.
Our second way to capture the spatial nature of a dangling spin degree of freedom
starts from the idea that it could be adiabatically connected to an uncoupled spin,
analogous to the Landau quasiparticle that is adiabatically connected to a free electron.
Thus, our program is to label each emergent spin-1/2 degree of freedom by 
a ``quasi spin'' operator $\mathbf{T}_{\alpha}$,
where the index $\alpha$ labels each region on the cluster with a local spin imbalance.
The $\mathbf{T}_{\alpha}$'s are idealized as having a spin-half algebra. 
The actual ``quasi spin'' excitation is a composite object 
involving multi-spin excitations, localized on a few sites.

Our assumption is that the quasi-spin quantum numbers are sufficient to label
all the QD states;  furthermore, we expect the action of any spin operator 
$\mathbf{S}_i$, when restricted to the QD states, practically reduces to a linear
combination of $\mathbf{T}_\alpha$'s acting on the quasi spins.
Specifically, let $\hat{P}_\QD$ be the projection operator onto the QD subspace. 
Let $\mathbf{S}_{i,\QD} \equiv \hat{P}_\QD \mathbf{S}_{i} \hat{P}_\QD$
Then, 
\begin{equation}
  \SIQD \cong \sum _\alpha u_{i}^{(\alpha)}\T_\alpha
\label{eq:ansatz}
\end{equation}
where each mode $u_{i}^{(\alpha)}$
has most of its weight on sites within the region
$\alpha$ and is expected to decay rapidly outside~\cite{FN-umode}.

Two operators $\hat{P},\hat{Q}$ are said to be \emph{orthogonal} when their Frobenius inner product 
$\left(\hat{P},\hat{Q}\right)_{F}\equiv{\rm Tr} \left(\hat{P}^{\dagger} \hat{Q} \right)$ is exactly zero. 
In this sense, the $\T_\alpha$ operators are \emph{orthogonal} to each other. Since each  $\T_\alpha$ 
is a quasi-spin 1/2 operator
its inner product with itself is 1/2 (for each spin component).

In light of Eq.\eqref{eq:ansatz}, we can also construct a good approximation $ \Ttilde_\alpha$ to each operator $\mathbf{T}_\alpha$, by
choosing any representative site $i$ in the region of $\alpha$ and normalizing
the restriction of its spin operator to the QD states:
  \begin{equation}
   \Ttilde_\alpha\equiv {\SIQD}/{(\sqrt{2}||\SIQD||)}
  \label{eq:Tconstruct}
  \end{equation}
where $||\hat{O}||\equiv (\hat{O},\hat{O})_{F}^{1/2}$
is the norm of any operator $\hat{O}$. Note that the $\Ttilde_\alpha$'s are
\emph{not} orthogonal to each other. A procedure to construct the  $\T_\alpha$'s from the $\Ttilde_\alpha$'s 
will be discussed later.

Given the proposed relationship of the bare spins to the quasi spins, we 
discuss two related but independent measurements to recover the mode
vectors $u^{(\alpha)}_{i}$ from numerically evaluated expectations.
%For all the measurements proposed here, it was sufficient to record (within DMRG) 
%all matrix elements of the form $\langle l'|S_i^{+}|l\rangle$ and energies $E_{l}$ 
%for eigenstates $l,l'$ in the QD subspace.
First, we consider the operator overlap $\omega_{ij}$ between two spins \emph{i} and \emph{j} on the lattice, 
defined to be,
\begin{equation}
\omega_{ij} \equiv \left( {S}_{i,\QD}^{+},{S}_{j,\QD}^{+} \right)_{F}
    \label{eq:omega_ij}
\end{equation}
We substitute our ansatz \eqref{eq:ansatz} into \eqref{eq:omega_ij} and use the operator orthogonality of the  $\mathbf{T}_\alpha$'s, to get 
$\omega_{ij}=\sum_\alpha u^{(\alpha)}_i u^{(\alpha)}_j$. If we consider a site \emph{i} to be well within a dangling region $\alpha$ (i.e. $ u^{(\alpha)}_i$ is relatively large)
then the amplitude on the remaining sites \emph{j} (but far away from other dangling regions) is approximately $u^{(\alpha)}_i u^{(\alpha)}_j$. 
Thus, the relative amplitudes of the mode vector can be recovered by this method. 

Our second measurement involves computation of the inter-site spin
susceptibility matrix, 
\begin{equation}
\chi_{ij}=\int_{0}^{\infty}{\langle\hat{S}_{i}^{z}(\tau)\hat{S}_{j}^{z}(0)\rangle}_{{GS}} d\tau=\sum_{n}\frac{\langle0|\hat{S}_{i}^{z}|n\rangle\langle n|\hat{S}_{j}^{z}|0\rangle}{E_{n}-E_{0}}
\label{eq:chiij}
\end{equation}
where $\tau$ is imaginary time, $|0\rangle$ denotes the ground state and $E_{n}$ is the energy of an excited state $|n\rangle$~\cite{FN-chi_ii}.

Though the sum runs over all excited states, it can be
well approximated by taking only the states in the QD subspace. 
Then $\chi_{ij}$ can also be expressed in terms of the mode profiles $u_{i}^{(\alpha)}$, 
\begin{equation}
\chi_{ij}=\sum _{\alpha \beta} u_i^{(\alpha)} u_j^{(\beta)} X_{\alpha\beta};
\qquad
X_{\alpha\beta} \equiv \sum_{n\in\mbox{QD}}
\frac{\langle0|\mathbf{T}_{\alpha}^{z}|n\rangle\langle n|\mathbf{T}_{\beta}^{z}|0\rangle} {E_{n}-E_{0}}
\label{eq:chiij-uiuj}
\end{equation}
Consider site $i$($j$) in dangling region $\alpha$ ($\beta$).
From Eq.~\eqref{eq:chiij-uiuj} it follows that $\chi_{ij} \approx u_i^{(\alpha)} u_j^{(\beta)} X_{\alpha\beta}$,
where the last factor is \emph{independent} of sites $i,j$ (so long as we stay within those
regions).  Within this approximation, the susceptibility matrix breaks up into blocks of rank 1 from which we can immediately pull out 
the $u_i^{(\alpha)}$ and $u_j^{(\beta)}$ modes.

\prlsec{Effective Hamiltonian in the Quasi degenerate subspace}
According to our ansatz \eqref{eq:ansatz}, there is a one-to-one correspondence between the QD Hilbert space 
and the Hilbert space of a set of abstract ``quasi spin'' operators $\mathbf{T}_\alpha$.
(For simplicity, assume they all have spin 1/2.)
The latter are labeled using an Ising basis $|\phi_t\rangle$,
where $t$ stands for the quantum numbers  $\{t_{1}^z,t_{2}^z,...t_{n_{d}}^z \}$,
with $t_\alpha=\pm 1/2$. We want to find the unitary matrix $\mathbf{M}$ of coefficients expressing the 
QD states $|l\rangle$ (in eigenenergy basis) in terms of the quasi-spin basis,
$|l\rangle = \sum _t M_{l t} |\phi_t\rangle$.

Using $\Ttilde_\alpha$ from \eqref{eq:Tconstruct}, we define
$\hat{Q}_\alpha^{\pm 1/2} \equiv (\frac{1}{2}\pm \Ttilde_\alpha^{z})$, which is
{\it almost}  a projection operator, and let $|\phitilde_t\rangle \propto \prod_{\alpha=1}^{n_{d}} \hat{Q}_\alpha^{t_\alpha} |n\rangle$ 
where $|n\rangle$ could be any QD state (that is not annihilated by the operator prefactors)
and $|\phitilde_t\rangle$ is normalized.  Finally, define a matrix $\underline{\Omega}$ by
$\Omega_{tt'}\equiv \langle \phitilde_t|\phitilde_{t'}\rangle$ -- which
is {\it almost} the identity matrix -- and construct the orthonormal
quasi spin basis of the QD states as $ |\phi_t\rangle \equiv \sum _{t'} (\underline{\Omega}^{-1/2})_{tt'} |\phitilde_{t'}\rangle$.  
The quasi spin operators $\mathbf{T}_\alpha$ are then defined so as to have the 
appropriate matrix elements in this basis.

Now consider the effective low energy Hamiltonian written in terms of the many body
eigenstates $|l\rangle$,
\begin{equation}
\Heff
\equiv \sum_{l \in \mbox{QD}}E_{l}|l\rangle\langle l|
= \sum_{tt'} h_{tt'} |\phi_t\rangle \langle \phi_{t'}|,
\label{eq:heff1}
\end{equation}
where $E_{l}$ is the eigenenergy of QD state $|l\rangle$,
and the matrix elements $h_{tt'}$  can be calculated since we know the
transformation between the bases $\{ |l\rangle \}$ and $\{ |\phi_t\rangle \}$.
Every term $|\phi_t\rangle \langle \phi_{t'}|$ can be uniquely expressed as a
polynomial in the spin operators $\{ T_\alpha^z \}$ and $\{ T_\alpha^\pm \}$.

The effective Hamiltonian \eqref{eq:heff1} then takes a new form:
\begin{equation}
\Heff\equiv \sum_{\mu,\nu}J_{\mu\nu}\mathbf{T}_{\mu}\cdot\mathbf{T}_{\nu}
		 + \mbox{multi spin terms}.
\label{eq:effham}
\end{equation}
(The two-spin terms must have this Heisenberg form due to the exact rotational symmetry retained
by the QD states.)

\begin{figure}[htpb]
\includegraphics[width=1\linewidth]{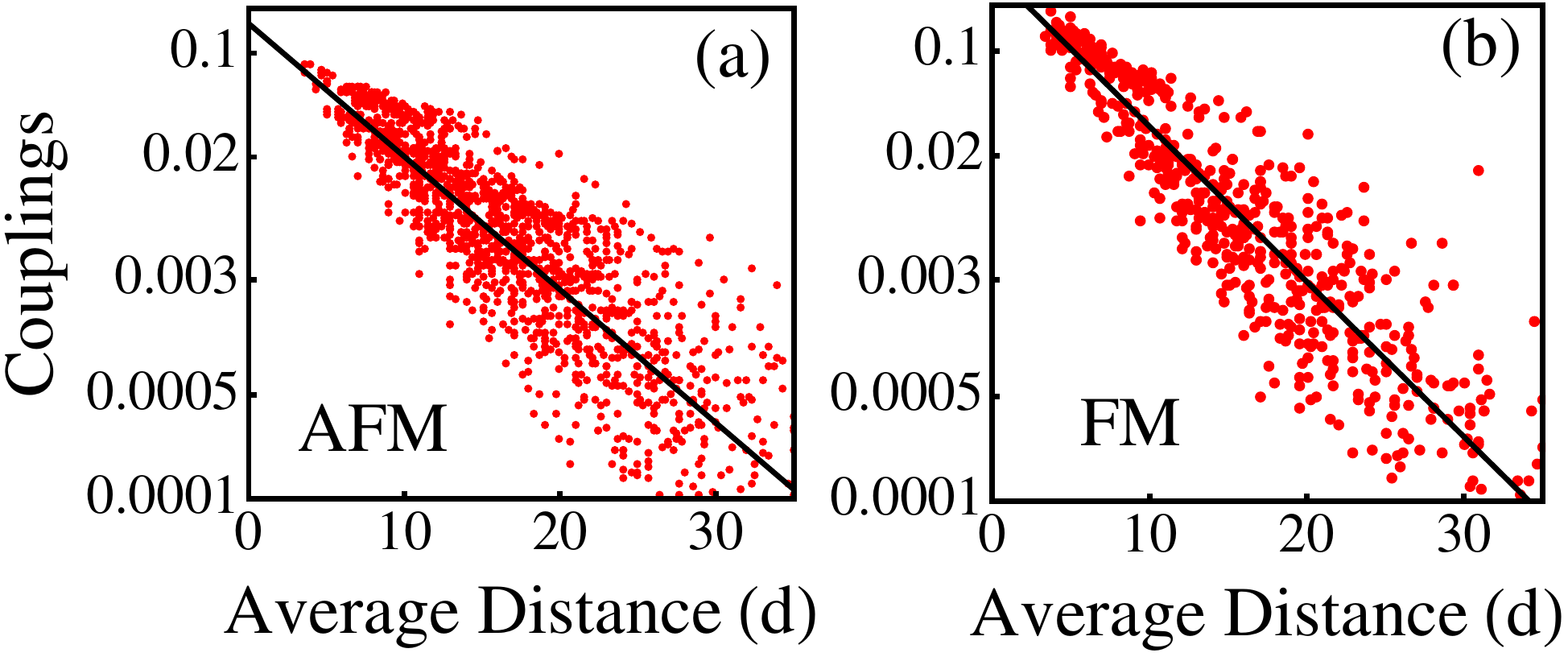}
\caption{(Color online) 
Effective couplings $J_{12}$ between two dangling spins as a function of their effective separation
obtained from clusters having exactly two dangling spins.
(a) dangling spins on opposite sublattices (antiferromagnetic coupling)
(b) on same sublattice (ferromagnetic coupling).
}
\label{fig:jafm}
\end{figure}

Although the magnitude of $J_{\mu\nu}$ depends on the detailed geometry of the cluster along
the path connecting dangling regions $\mu$ and $\nu$, roughly speaking it decays exponentially
with distance (using as metric the number of steps within the Bethe lattice, the so-called
"chemical" distance).  This is quantified by the scatter plots in 
Fig.~\ref{fig:jafm}, for an ensemble limited to clusters of equal size 
$N=100$ \cite{FN-size_coupling} each having two dangling spin-1/2 spins 
($n_{d}=\nhalf=2$).
Since each dangling region is, in general, spread out over multiple sites,
we must define an ``effective distance'' $ \bar{d}_{\mu\nu} \equiv \sum_{ij}  |u^{(\mu)}_{i}|^2 |u^{(\nu)}_{j}|^2 d_{ij}$  between two of them, where 
$d_{ij}$ is the distance between sites $i$ and $j$ belonging to dangling regions $\mu$ and $\nu$,
and the amplitudes $u^{(\alpha)}_{i}$ for mode $\alpha$ are normalized.

Fig.~\ref{fig:jafm} shows that indeed $J_{\mu\nu} \approx J_0 e^{-\bar {d}_{\mu\nu}/\xi}$, 
where $(J_0,\xi) \approx (+0.18(3),4.64(2))$ for an even/odd pair of dangling spins, 
which are {\it always} antiferromagnetically coupled, or $(-0.33(2),4.61(3))$ for a pair
on the same sublattice. In the \emph{ferromagnetic} case, choosing to fit to clusters which do not form 
a spin-1 gives parameter values 
closer to the \emph{antiferromagnetic} case.

We considered another ensemble (not plotted) of clusters with $N=50$ having four dangling spins 
($n_{d}=4$) and obtained the effective Hamiltonian using the same prescription. 
In it, we found, the non-pair terms in \eqref{eq:effham}
typically account for a weight of at most 5\% (using the 
Frobenius norm),
confirming that the effective Hamiltonian is well approximated
by pairwise Heisenberg exchange (at least in the limit of dilute monomer concentration). 

\prlsec{Conclusion} 
The spin 1/2 Heisenberg antiferromagnet on
Bethe lattice percolation clusters has (composite)
low-energy degrees of freedom with the quantum numbers
of a spin, arising
wherever there is a local imbalance between the even and odd sublattices~\cite{wangandsandvik},
(A similar imbalance determines the low energy spectra of {\it regular} Cayley trees~\cite{pure_bethe}.) 
Each of these emergent ``dangling spins'' is associated
with a profile ($u_i^{(\alpha)}$ in the text) that plays the
role of a ``spinon wavepacket wavefunction''~\cite{sandvik-spinon}. 
We leave to a future publication~\cite{whythelowering}
the fundamental reason {\it why} a dangling spin decouples from the rest of the cluster.

Our picture of "dangling spins" can be tested experimentally
using local probes.  For example, NMR at a temperature
scale between the mediated interactions in~\eqref{eq:effham} and the
bare interactions in~\eqref{eq:hamiltonian}, in the presence of a field,
gets a line shape mirroring the dangling-spin profile $u_i$,
while zero-field muon spin resonance can detect the absence
of an order parameter on sites away from the dangling spins.

The dangling spins interact via small, unfrustrated effective 
Heisenberg couplings.  
If one adopts the fitted exponential decay from Fig.~\ref{fig:jafm} as our
{\it definition} of the effective Hamiltonian, it should be possible to
study clusters with thousands of sites and finally explain
the scaling of the spin gap $\Delta_{low}$ with cluster size
found in Refs.~\onlinecite{wangandsandvik},
by use of the strong disorder renormalization group 
method~\cite{BhattLee}.

Finally, we found that spin correlations decay exponentially in balanced regions, 
which are dimerized, but revived on the dangling spins.
This suggests locally unbalanced regions may be crucial for the propagation of long range
antiferromagnetic order on percolation clusters.

\prlsec{Acknowledgement} HJC would like to thank Garnet Chan for discussions
on the DMRG technique. We also thank Anders Sandvik for discussions
and Andreas L\"auchli for pointing out references on related work. We
acknowledge support from National Science Foundation grant NSF DMR-1005466. SP was also supported by the Indo-French Center for the Promotion of Advanced Research (IFCPAR/CEFIPRA) under Project 4504-1 during a part of this work.


\begin{thebibliography}{10}
\bibitem{FN-balanced}{Balanced clusters necessarily have a singlet ground state.}
\bibitem{FN-sixdangling}{Cases of six dangling spins could be qualitatively understood by looking at approximate energies 
obtained at the end of the initialization step of the DMRG (as described in~\cite{pure_bethe}).}
\bibitem{FN-QD} Interactions between dangling spins separated by different 
length scales create a hierarchy of energy levels. However, there will always 
be a set of low lying states arising from splittings between the most distant spins on the cluster. 
We label these states as 'QD' spectrum for computing $r$.
\bibitem{FN-spin1} This prescription for predicting spin ones was found to be invalid for 
~3\% clusters with $n_{d}=4$. A more stringent criterion relies on the ratio of the distance 
at which two momomers couple to form a spin-1 and the separation of this unit from all other spins.
\bibitem{FN-umode} Large fluctuations are associated with local imbalance in 
linear spin-wave theory~\cite{Harris_Kirkpatrick}, but our work~\cite{whythelowering} shows this is insufficient to 
explain our exponentially localized degrees of freedom emerging in a background which seems to lack 
N\'eel order.
\bibitem{FN-size_coupling} 
We observed a weak dependence of the coupling 
(for a given "effective distance") on the size of the cluster.
\end{thebibliography}
\end{document}